# DISTRIBUTED FIREWALLS AND IDS INTEROPERABILITY CHECKING BASED ON A FORMAL APPROACH

Kamel Karoui[1] , Fakher Ben Ftima[2] and Henda Ben Ghezala[3]

[1]RIADI, ENSI, University of Manouba, Manouba, Tunisia
`kamel.karoui@insat.rnu.tn`

[2]RIADI, ENSI, University of Manouba, Manouba, Tunisia
`fakher.benftima@infcom.rnu.tn`

[3]RIADI, ENSI, University of Manouba, Manouba, Tunisia
`Henda.bg@cck.rnu.tn`

*ABSTRACT*

*To supervise and guarantee a network security, the administrator uses different security components, such as firewalls, IDS and IPS. For a perfect interoperability between these components, they must be configured properly to avoid misconfiguration between them. Nevertheless, the existence of a set of anomalies between filtering rules and alerting rules, particularly in distributed multi-component architectures is very likely to degrade the network security. The main objective of this paper is to check if a set of security components are interoperable. A case study using a firewall and an IDS as examples will illustrate the usefulness of our approach.*

## 1. INTRODUCTION

Security components are crucial elements for a network security. They have been widely deployed to secure networks. A security component is placed in strategic points of a network, so that, all incoming and outgoing packets have to go through it [1, 2].

Generally, to enhance and guarantee the system safety, the administrator enforces the network security by distributing many security components over the network. This implies cohesion of the security functions supplied by these components. A misconfiguration or a conflict between security components set of rules means that a security component , may either accept some malicious packets, which consequently creates security holes , or discard some legitimate packets, which consequently disrupts normal traffic. Both cases could cause irreparable consequences [3, 4].

Unfortunately, it has been observed that most security components are poorly designed and have many anomalies which implies many repercussions, both on their functioning and on their interoperability, with other security components. Given the importance of the network security, such errors are not acceptable [5, 6].

Considering the impact of the poorly designed security component set of rules on the networks' safety, it is necessary to[7, 8]:

- Specify and check its set of rules correctness before its installation in a network.
- Verify the security component interoperability with other security components on the network.

Several models are proposed for security components analysis[9,10,11].In our work, we propose a decision tree-based approach composed of three processes. In the first one, we verify and correct misconfigurations in the security component set of rules and generate a new set free of anomalies. In the second one, we will check the interoperability between several security components in the network. If the interoperability between distributed security components in a network is not confirmed, we will apply a correction process which applies a formal model to guarantee the security components interoperability.

The remaining parts of the paper are organized as follows; section 2 presents the proposed approach. Section 3 presents the security component set of rules extraction, verification and correction process steps. Section 4 presents the security components interoperability checking process steps. Section 5 presents the interoperability correction process steps and section 6 concludes the paper.

## 2. THE PROPOSED APPROACH

In order to verify the security components interoperability in distributed architectures, we propose an approach composed of the following processes (see figure 1):

**Initial process: Security components positioning checking**
There are several types of security components; filtering components and alerting ones. So, to guarantee the network security, the administrator installs, generally, filtering security components in strategic points (for example internet or traffic flowing from an external network). Then, as a complementary equipment of defense, the administrator installs an alerting security component. Therefore, the security components' positioning is very important. Inversing this order creates security holes that will allow malicious traffic to sneak into the network. Thus, the following processes are applicable to filtering/filtering, filtering/alerting or alerting/alerting security components.

**Process 1: Security components set of rules extraction, verification and correction**
This process is composed of the following steps:
- Step A: Extraction of the security component set of rules
- Step B: Formal security component set of rules checking
- Step C: Formal security component set of rules correction

**Process 2: Security components interoperability checking**
This process is executed for a set of security components two by two. It is composed of the following steps:
- Step D: Security components set of attributes extraction
- Step E: Security components set of rules extension
- Step F: Formal security components interoperability checking

**Process 3: Security components interoperability correction**
This process is executed once the security components interoperability is not confirmed (see step F). It is composed of the following steps:

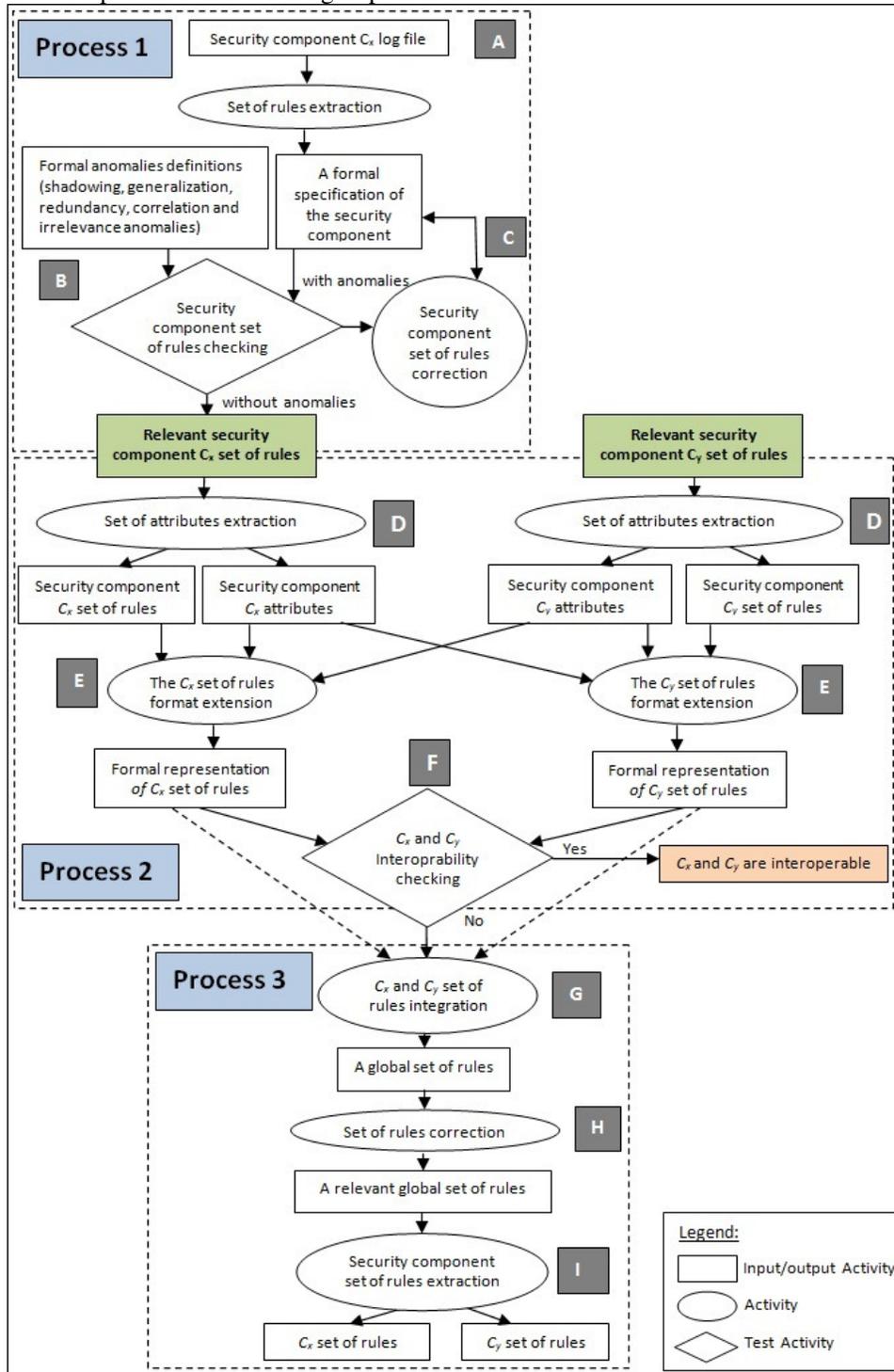

Figure 1.The proposed interoperability checking approach

# 3. PROCESS 1: SECURITY COMPONENTS SET OF RULES EXTRACTION, VERIFICATION AND CORRECTION

In this section, we will tack in details the security component set of rules extraction, verification and correction process (see process 1 in figure 1). This process aims to:
- represent the security component set of rules into a standardized format,
- verify the security component set of rules relevancy,
- correct the security component set of rules incoherencies.

These steps prepare the security component for interoperability verification with the specific security policy.

## 3.1 Step A: Extraction of The Security Component Set of Rules (See Figure 1)

In previous works [12], we have shown how to extract a security component set of rules from a security component log file. In the followings sub-sections, we will define the security component set of rules format. This format will be represented by a decision tree approach for anomalies detection and correction in the next sections.

### 3.1.1 Formal Security Component Set of Rules Representation

For a security component $C_x$, having a set of $t$ rules $R_x = \{r_1, r_2, \ldots r_i, \ldots r_t\}$, each rule is defined formally over a set of $n$ attributes $A_1, A_2, \ldots, A_n$. $A_n$ is a specific attribute called decision attribute. We define a general rule format as follows:

$$r_i : [e_{1,i} \wedge e_{2,i} \wedge \ldots e_{j,i} \wedge \ldots \wedge e_{n-1,i}] \rightarrow e_{n,i}$$

where:

- $e_{j,i}$ with $1 \leq j \leq n-1$ is the value of attribute $A_j$ in the rule $r_i$. It can be a single value (for example: *UDP*, *80*,...) or a range of values (for example: *[192.120.30.30/24, 192.120.30.50/24]*). $e_{j,i} \subseteq D_j$ where $D_j$ is the domain of the attribute $A_j$ with $1 \leq j \leq n-1$. For instance, for an attribute $A_1 = $ "protocol", its attribute domain is $D_1 = \{UDP, TCP, ICMP\}$ and $e_{1,1} = $ "UDP".

- $[e_{1,i} \wedge e_{2,i} \wedge \ldots e_{j,i} \wedge \ldots \wedge e_{n-1,i}]$ is the conjunctive set of the rule $r_i$ attributes values with $1 \leq j \leq n-1$.

- $e_{n,i}$ is a specific value of the attribute $A_n$. It takes its value from the set of values *{accept, deny, discard, pass}*.

**Example 1:**

Let's take a security component $C_x$. If we suppose $A_1$ and $A_2$, respectively, the attributes "*source address*" and "*destination address*", and if we suppose that $D_1$: *[192.120.*.*/24]*, $D_2$: *[128.160.*.*/24]* and $D_3 = $ *{accept, deny}*, we can define a rule $r_i$ with IP addresses *[192.120.30.*/24]* that accepts access for hosts belonging to $D_1$ to access to hosts with IP addresses *[128.160.40.*/24]* belonging to $D_2$, as follows:

$$r_i : 192.120.30.*/24 \wedge 128.160.40.*/24 \rightarrow accept$$

We can define the following properties:

**Property 1:**

Let's take an IP packet $P$ and $p_1, p_2, \ldots p_m$ the packet header fields (with $1 \leq m \leq n-1$). We say that the IP packet $P$ verifies a rule $r_i : [e_{1,i} \wedge e_{2,i} \wedge \ldots e_{j,i} \wedge \ldots \wedge e_{n-1,i}] \rightarrow e_{n,i}$ in the security component $C_x$, if $p_1 \in e_{1,i} \wedge p_2 \in e_{2,i} \wedge \ldots \wedge p_m \in e_{m,i}$.

For example, the IP packet $P : ([192.120.30.5/24] \wedge [128.160.40.25/24])$ verifies the rule $r_i$ (see Example 1).

**Property 2:**
We say that a security component $C_x$ with $t$ rules $\{r_1, r_2 \ldots r_t\}$ is reliable if, for any IP packet $P$, there exists one rule $r_i$ in $C_x$ ($1 \leq i \leq t$) that verifies the packet (see property 1).

### 3.1.2 The Decision Tree Approach

We propose to use the decision tree model to describe a security component $C_x$ set of rules. A decision tree is a formal representation defined by 3 types of entities (see figure 2):
- Nodes: represent the different attributes of a rule. They are schematized by labeled rectangles in the tree.
- Edges: connect the decision tree nodes. They are labeled by values or a range of values taken from the parent node domain. They are schematized by a labeled and directed arrow from the parent node to the outgoing nodes.
- Leaves: are terminal nodes representing the path identification. They are schematized by a labeled circle in the tree.

We can represent a security component $C_x$ with $t$ rules by a decision tree where each path from the root to a terminal node represents a rule of $C_x$. Those paths are called **branches** $b_i$ with $1 \leq i \leq t$. So, a decision tree $DT$ with the attributes $A_1, A_2, \ldots A_n$ is a tree that satisfies the following conditions:
- The root of the tree representing the attribute $A_1$ is labeled by $A_{1,w}$ where $w$ represents the branch $b_w$ in the decision tree $DT$ with $1 \leq w \leq t$ and $t$ represents the number of $DT$ branches.
- For example, in figure 2, the root is labeled by $A_{1,1} = A_{1,2} = A_{1,3} = A_{1,4} = \ldots = A_{1,t} =$ "Protocol".
- Each non-terminal node representing the attribute $A_m$, is denoted $A_{m,w}$ where $m$ ($1 \leq m \leq n$) represents its level in the tree and $w$ ($1 \leq w \leq t$) the belonging of the node to the branch $b_w$.
- For example, in figure 2, nodes in the second level are labeled by $A_{2,1} = A_{2,2} = A_{2,3} = A_{2,4} =$ "Source port".
- Each edge, connecting two nodes $A_m$ and $A_{m+1}$ which represents the attribute $A_m$ value is denoted $e_{m,w}$ where $m$ ($1 \leq m \leq n$) represents the level in the tree and $w$ ($1 \leq w \leq t$) the branch in the tree.
- For example, in figure 2, we note *"UDP"* and *"ICMP"* the labeled edges connecting the attributes *"Protocol"* and *"Source port"*.
- Each terminal node is labeled with the specific value *"Null"*. It represents the termination of a branch in the decision tree $DT$.
- Each path in the decision tree from the root to the leaf is identified by the branch identification $b_w$ ($1 \leq w \leq t$) (see figure 2).

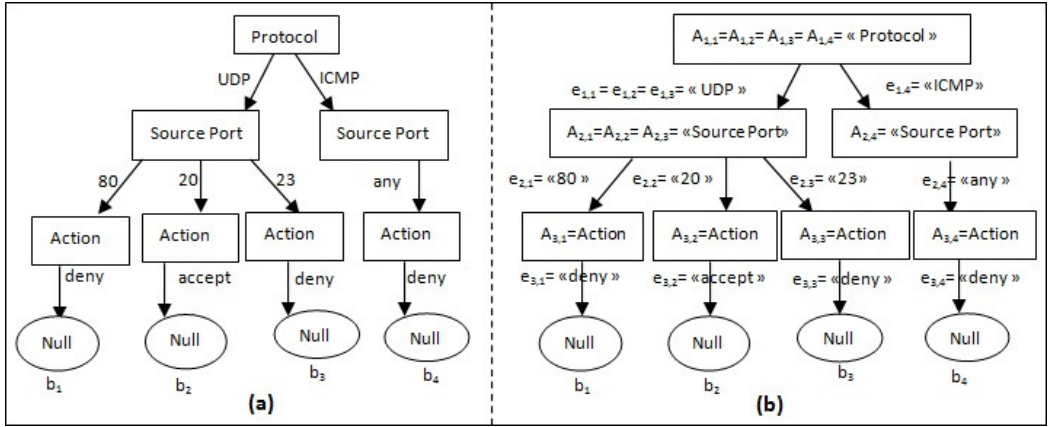

Figure 2. A decision tree representation

We define the set of labeled edges belonging to a level *m* in *DT* as follows:
$e_m = \{e_{m,1}, e_{m,2}, \ldots e_{m,t}\}$. We note that $e_m$ is a sub-set of $D_m$ (domain of $A_m$). For example, in figure 2, the set of labeled edges belonging to the level 2 is: $e_2 = \{e_{2,1}=\text{«80»}, e_{2,2}=\text{«20»}, e_{2,3}=\text{«23»}$ and $e_{2,4}=\text{«any»}\}$

A rule $r_w$ is represented in a decision tree *DT* by a branch $b_w$ as follows:
  $b_w: A_{1,w}\text{-}e_{1,w}\text{-}A_{2,w}\text{-}e_{2,w}\ldots A_{m,w}\text{-}e_{m,w} \ldots\ldots A_{n,w}\text{-}e_{n,w}\text{-}Null$ with $1 \leq m \leq n$ and $1 \leq w \leq t$

Based on (1), we define the suffix_node of $b_w$ as follows:
  $Suffix\_node(b_w, A_{i,w}) = A_{i,w}\text{-}e_{i,w}\ldots A_{m,w}\text{-}e_{m,w} \ldots\ldots A_{n,w}\text{-}e_{n,w}\text{-}Null$
with $1 \leq i \leq n-1$ and $1 \leq w \leq t$. The $Suffix\_node(b_w, A_{i,w})$ function returns the postfix of $b_w$ from the node $A_{i,w}$.

Also, based on (1), we define the suffix_edge of $b_w$ as follows:
  $Suffix\_edge(b_w, e_{i,w}) = e_{i,w}\ldots A_{m,w}\text{-}e_{m,w} \ldots\ldots A_{n,w}\text{-}e_{n,w}\text{-}Null$ with $1 \leq i \leq n-1$ and $1 \leq w \leq t$
The $Suffix\_edge(b_w, e_{i,w})$ function returns the postfix of $b_w$ from the edge $e_{i,w}$.

A branch $b_w$ in the decision tree *DT* represents the rule $r_w$ as follows:
$r_w: e_{1,w} \wedge e_{2,w} \wedge \ldots e_{n-1,w} \rightarrow e_{n,w}$ with $e_{1,w} \subseteq D_1, e_{2,w} \subseteq D_2, \ldots e_{n-1,w} \subseteq D_{n-1}, e_{n,w} \subseteq D_n$

For example, in figure 2, the branch
$b_2: A_{1,2}=$ "Protocol"- $e_{1,2}=$"UDP"- $A_{2,2}=$"Source port" - $e_{2,2}=$ "20"- $A_{3,2}=$"Action "- $e_{3,2}=$ "Accept" -Null
represents the following rule: $UDP \wedge 20 \rightarrow accept$
with $UDP \in$ "Protocol", $20 \in$ "Source port", $accept \in$ "Action"

In figure 2, we note that in the decision tree *DT*, $b_2$ and $b_3$ have the same prefixes; the attributes $A_{1,1} = A_{1,2} =$"Protocol" and $A_{2,1} = A_{2,2} =$"Source port". Also, the labeled edges $e_{1,1}= e_{1,2} =$"TCP". This is due to the fact that they share, respectively, the same node and the same branch. So, $b_2$ can also be written as follows:
$b_2: A_{1,1}=$ "Protocol"- $e_{1,1}=$"UDP"- $A_{2,1}=$"Source port" - $e_{2,2}=$ "20"- $A_{3,2}=$"Action "- $e_{3,2}=$ "Accept"-Null

### 3.1.3 Case study: Security component set of rules extraction

Let's take a firewall *FW* as security component. By applying the set of rules extraction process on *FW* log file [12], we obtain the following set of rules (see step A in figure 3).

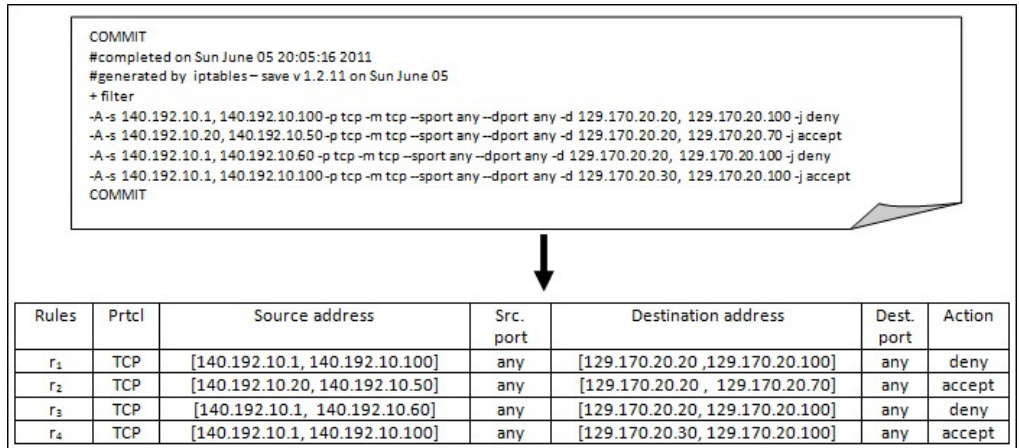

Figure 3. The firewall *FW* set of extracted rules $R_{FW}$

Using the decision tree approach described previously (see section 3.1.2), we represent the firewall *FW* set of rules $R_{FW}$ by a decision tree $DT_{FW}$ as follows (see figure 4):

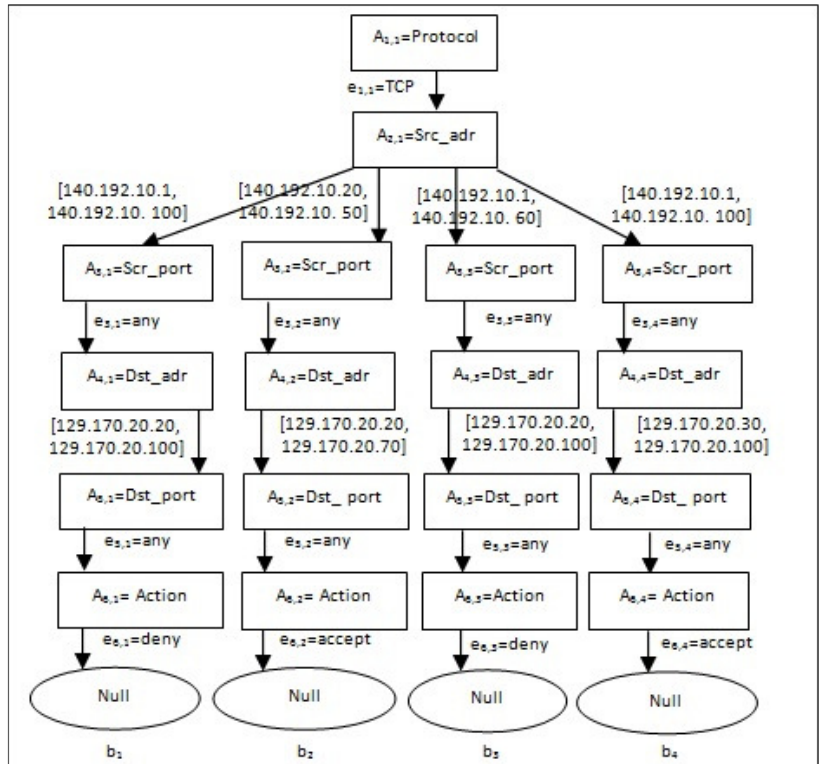

Figure 4. $DT_{FW}$: The firewall *FW* decision tree

where:

$b_1$: "TCP-[140.192.10.1,140.192.10.100]-any-[129.170.20.20,129.170.20.100]-any-deny-Null" in $DT_{FW}$
corresponds to the rule $r_1$ in the set of rules $R_{FW}$.
$b_2$: "TCP- [140.192.10.20,140.192.10.50]-any-[129.170.20.20,129.170.20.70]-any-accept-Null" in $DT_{FW}$
corresponds to the rule $r_2$ in the set of rules $R_{FW}$.

$b_3$ : "TCP- [140.192.10.1,140.192.10.60]-any-[129.170.20.20,129.170.20.100]-any-deny-Null" in $DT_{FW}$
corresponds to the rule $r_3$ in the set of rules $R_{FW}$.
$b_4$ :"TCP-[140.192.10.1,140.192.10.100]-any-[129.170.20.30,129.170.20.100]-any-accept-Null"in $DT_{FW}$ corresponds to the rule $r_4$ in the set of rules $R_{FW}$.

## 3.2 Step B: Formal Security Component Set Of Rules Checking (See Figure 1)

Let's take a security component $C_x$. In order to study the security component $C_x$ set of rules correctness, we have chosen to represent it by a decision tree. This representation will allow us to have a better illustration of the security component. Several works [13,14,15] have defined a set of anomalies detectable between rules in a security component called "component anomalies".In the following, we will study these anomalies using the decision tree formalism. Then, we will propose a formal method to remove them.

### 3.2.1 Formalization Of Relations Between Rules

Let's take a decision tree $DT$ composed of $t$ branches (representing $t$ rules). As mentioned above (see section 3.1.2), a branch $b_i$ corresponding to a rule $r_i$ in $DT$ is formalized as follows:

$$b_i : A_{1,i}e_{1,i}....A_{m,i}e_{m,i}....A_{n,i}e_{n,i}Null \quad with\ 1 \le i \le t\ and\ 1 \le m \le n$$

In [13,14], the authors have defined the followings definitions:

**Definition1:** Rules $r_i$ and $r_j$ are *exactly matching* if every field in $r_i$ is equal to its corresponding field in $r_j$. Formally, $r_i \Re_{EM} r_j$ if $\forall 1 \le m \le n-1,\ r_j[A_m] = r_i[A_m]$ with $1 \le i < j \le t$.

In the same way, we define that branches $b_i$ and $b_j$ are *exactly matching* if every labeled edge in $b_i$ is equal to its corresponding labeled edge in $b_j$. Formally, $b_i \Re_{EM} b_j$ if $\forall 1 \le m \le n-1,\ e_{m,j} = e_{m,i}$ with $1 \le i < j \le t$.

**Definition2:** Rules $r_i$ and $r_j$ are *inclusively matching* if they do not exactly match and if every field in $r_i$ is a subset or equal to its corresponding field in $r_j$. $r_i$ is called the subset while $r_j$ is called the superset.
Formally, $r_i \Re_{IM} r_j$ if $\forall 1 \le m \le n-1,\ r_i[A_m] \subseteq r_j[A_m]$ with $1 \le i < j \le t$.

In the same way, we define that branches $b_i$ and $b_j$ are *inclusively matching* if they do not exactly match and if every labeled edge in $b_i$ is a subset or equal to its corresponding labeled edge in $b_j$. $b_i$ is called the subset while $b_j$ is called the superset. Formally, $b_i \Re_{IM} b_j$ if $\forall 1 \le m \le n-1,\ e_{m,i} \subseteq e_{m,j}$ with $1 \le i < j \le t$

**Definition3:** Rules $r_i$ and $r_j$ are *correlated* if some fields in $r_i$ are subsets or equal to the corresponding fields in $r_j$, and the rest of the fields in $r_i$ are supersets of its corresponding fields in $r_j$. Formally, $r_i \Re_c r_j$ if

$$\forall 1 \le m \le n-1, (r_i[A_m] \subset r_j[A_m]) \vee (r_i[A_m] \supset r_j[A_m]) \vee (r_i[A_m] = r_j[A_m])\ with\ 1 \le i < j \le t$$

In the same way, we define that branches $b_i$ and $b_j$ are *correlated* if some labeled edges in $b_i$ are subsets or equal to its corresponding labeled edges in $b_j$, and the rest of the labeled edges in $b_i$ are supersets of the corresponding labeled edges in $b_j$. Formally, $b_i \Re_c b_j$ if

$$\forall 1 \le m \le n-1, (e_{m,i} \subset e_{m,j}) \vee (e_{m,i} \supset e_{m,j}) \vee (e_{m,i} = e_{m,j})\ with\ 1 \le i < j \le t$$

**Definition4:** Rules $r_i$ and $r_j$ are *disjoints* if there exist at least one field in $r_i$ different from its corresponding field in $r_j$. Formally, $r_i \mathfrak{R}_D r_j$ if $\exists\, 1 \leq m \leq n-1$, $(r_i[A_m] \neq r_j[A_m])$ with $1 \leq i < j \leq t$

In the same way, we define that branches $b_i$ and $b_j$ are *disjoints* if there exist at least a labeled edge in $b_i$ different from its corresponding labeled edge in $b_j$. Formally, $b_i \mathfrak{R}_D b_j$ if

$\exists\, 1 \leq m \leq n-1, (e_{m,i} \neq e_{m,j})$ with $1 \leq i < j \leq t$

### 3.2.2 Security Components Anomalies Detection

An anomaly in a security component is the result of the following cases [16,17]:
- The existence of two or more rules that may match the same packet
- The existence of a rule that can never match any packet on the network paths that cross the security component.

In the following, we classify different anomalies that may exist between rules in a security component.

**Property 3: Shadowing Anomaly**
In a set of rules $R$, a rule $r_j$ is shadowed by a previous rule $r_i$ when $r_i$ matches all the packets that match $r_j$, such that the shadowed rule $r_j$ will never be activated. In a decision tree $DT$, for any two branches $b_i$ and $b_j$ with $1 \leq i < j \leq t$, $b_j$ is shadowed by $b_i$ if and only if, $(b_j \mathfrak{R}_{IM} b_i) \wedge (e_{n,j} \neq e_{n,i})$

**Property 4: Generalization Anomaly**
The generalization anomaly is the reverse of the shadowing anomaly i.e. in a set of rules $R$, a rule $r_j$ is a generalization of a preceding rule $r_i$ if, on the one hand, the rule $r_j$ can match all the packets that match the rule $r_j$ and, on the other hand, the two rules have different actions.
In a decision tree $DT$, for any two branches $b_i$ and $b_j$ with $1 \leq i < j \leq t$, $b_j$ is a generalization of $b_i$ if and only if, $(b_i \mathfrak{R}_{IM} b_j) \wedge (e_{n,i} \neq e_{n,j})$

**Property 5: Redundancy Anomaly**
In a set of rules $R$, a rule $r_j$ is redundant to a rule $r_i$ if $r_j$ performs the same action on the same packets as $r_i$. In the way, if the redundant rule $r_j$ is removed, the safety of the security component will not be affected. In a decision tree $DT$, for any two branches $b_i$ and $b_j$ with $1 \leq i < j \leq t$, $b_j$ is redundant to $b_i$ if and only if, $(b_j \mathfrak{R}_{IM} b_i) \wedge (e_{n,i} = e_{n,j})$

**Property 6: Correlation Anomaly**
In a set of rules $R$, two rules $r_j$ and $r_i$ are correlated if, on the one hand, the first rule $r_j$ matches some packets that match the second rule $r_i$, and the second rule $r_i$ matches some packets that match the first rule $r_j$ and, on the other hand, the two rules have different actions.
In a decision tree $DT$, for any two branches $b_i$ and $b_j$ with $1 \leq i < j \leq t$, $b_j$ and $b_i$ are correlated if and only if, $(b_i \mathfrak{R}_c b_j) \wedge (e_{n,i} \neq e_{n,j})$

### 3.2.3 Case study: Security Component Anomalies Detection
In our case study, in figure 4, we note that the firewall *FW* contains some misconfigurations between its rules:

- $b_2$ (representing $r_2$) is shadowed by $b_1$ (representing $r_1$). More precisely:
$(e_{1,2} = e_{1,1}) \wedge (e_{2,2} \subset e_{2,1}) \wedge (e_{3,2} = e_{3,1}) \wedge (e_{4,2} \subset e_{4,1}) \wedge (e_{5,2} = e_{5,1}) \wedge (e_{6,2} \neq e_{6,1})$

Also, in the same figure, $b_4$ (representing $r_4$) is shadowed by $b_1$ (representing $r_1$). More precisely: $(e_{1,4} = e_{1,1}) \wedge (e_{2,4} = e_{2,1}) \wedge (e_{3,4} = e_{3,,1}) \wedge (e_{4,4} \subset e_{4,1}) \wedge (e_{5,4} = e_{5,1}) \wedge (e_{6,4} \neq e_{6,1})$

- $b_3$ (representing $r_3$) is a generalization of $b_2$ (representing $r_2$). More precisely:
$(e_{1,2} = e_{1,3}) \wedge (e_{2,2} \subset e_{2,3}) \wedge (e_{3,2} = e_{3,3}) \wedge (e_{4,2} = e_{4,3}) \wedge (e_{5,2} = e_{5,3}) \wedge (e_{6,2} \neq e_{6,3})$

- $b_3$ (representing $r_3$) is redundant to $b_1$ (representing $r_1$). More precisely:

$$(e_{1,3} = e_{1,1}) \wedge (e_{2,3} \subset e_{2,1}) \wedge (e_{3,3} = e_{3,1}) \wedge (e_{4,3} = e_{4,1}) \wedge (e_{5,1} = e_{5,3}) \wedge (e_{6,3} = e_{6,1})$$

- $b_3$ (representing $r_3$) is correlated to $b_4$ (representing $r_4$). More precisely:

$$(e_{1,3} = e_{1,4}) \wedge (e_{2,3} \subset e_{2,4}) \wedge (e_{3,3} = e_{3,4}) \wedge (e_{4,3} \supset e_{4,4}) \wedge (e_{5,3} = e_{5,4}) \wedge (e_{6,3} \neq e_{6,4})$$

In the next section, we will present a new approach to remove them.

## 3.3 Step C: Formal security component set of rules correction (see figure 1)

By studying the previous anomalies properties on the decision tree (see section 3.2.2), we propose a fundamental property guarantying that the decision tree is free of anomalies. We call this property the "relevancy property".

**Property 7: Relevancy**
let's take a decision tree $DT$ with $t$ branches. $G(A_{m,w}) = \{e_{m,i}, ...., e_{m,i+k}\}$ is the set of all $k$ ($k>1$) outgoing labeled edges from the node $A_{m,w}$ with $1 \leq i \leq t-k$. The decision tree $DT$ is relevant, if and only if for any two edges $e_{m,i}$ and $e_{m,j}$ belonging to $G(A_{m,w})$, we have:

$$\left[ e_{m,i} \cap e_{m,j} = \phi \right] \wedge \left[ e_{k,i} \in D_k \wedge e_{z,i} \in D_z \right]$$

where k represents the "source address" attribute and z represents the "destination address" attribute.

For example, in figure 2, the node $A_{1,1}$ (noted also *"protocol"*) has two outgoing edges labeled $e_{1,1} = $ *"UDP"* and $e_{1,4} = $ *"ICMP"*. Thus, $G(A_{1,1}) = \{e_{1,1}, e_{1,4}\}$. We note that $e_{1,1} \cap e_{1,4} = \phi$

We can prove that a decision tree verifying the relevancy property (property 7) is free of the anomalies presented above (see properties 3 to 6 in section 3.2.2).

**Lemma 1:**

A decision tree $DT$ verifying the relevancy property doesn't contain the previous anomalies (i.e. the shadowing anomaly, the generalization anomaly, the correlation anomaly and the redundancy anomaly) (see section 3.2.2).

For example, the decision tree of figure 4 is non-relevant because branches $b_1$ and $b_2$ verify the shadowing anomaly.
A similar reasoning on properties 4, 5 and 6 proves that a decision tree with the generalization anomaly, the redundancy anomaly and the correlation anomaly is a non-relevant decision tree.

To remove the decision tree $DT$ misconfigurations, we will build another decision tree called **R**elevant **D**ecision **T**ree (*RDT*) which verifies the relevancy property (see property 7). The proposed *RDT* will be presented in the next section.

### 3.3.1 The Relevant decision Tree (RDT):

In the following sub-sections, we first start by explaining the *RDT* construction principle informally, then we will present it with a formal algorithm. To do that, we need to take into account some assumptions:

**Assumption 1:**
In a security component $C_x$ with a set of $t$ rules $R_x$ $(r_1, r_2, ....r_i, ...r_t)$, if a rule $r_i$ is applicable for an IP paquet, so the remaining set of rules i.e rules from $r_{i+1}$ to $r_t$ is ignored. This assumption preserve the set of rules order during the *RDT* construction algorithm treatement.

**Assumption 2:**
In a security component $C_x$ with a set of $t$ rules $R_x$ $(r_1, r_2, .... r_i, ... r_t)$, if there are anomalies between $r_i$ and $r_{i+1}$, $r_{i+1}$ will be corrected according to $r_i$. This assuption ensure that all rules in a security component have the same importance.

### 3.3.2 The RDT construction

In this section, we will take some examples to explain the principle of decision tree branches' construction. The decision tree construction will be done recursively and will be explained in the decision tree construction algorithm (see section 3.3.3).

Let's take a security component $C_x$ with a set of 2 rules $R_x \{r_1, r_2\}$ having the following format:
$$r_1 : e_{1,1} \wedge e_{2,1} \rightarrow e_{3,1} \text{ with } e_{1,1} \subseteq D_1, e_{2,1} \subseteq D_2, e_{3,1} \subseteq D_3$$
$$r_2 : e_{1,2} \wedge e_{2,2} \rightarrow e_{3,2} \text{ with } e_{1,2} \subseteq D_1, e_{2,2} \subseteq D_2, e_{3,2} \subseteq D_3$$

where $D_1=[1-30]$ represents the "source address" (Src adr) domain, $D_2$ represents the "destination address" (Dest adr) domain and $D_3$ represents the "Action" (Action) domain. The RDT construction algorithm builds the first branch $b_1$ (representing the first rule $r_1$) of the tree, and then joins $b_2$ (representing *the rule $r_2$*) to the tree. There are several cases to study:

**-Case 1**: $r_1 \Re_D r_2$ : Let's take, for example, the set of 2 rules $R_x \{r_1, r_2\}$ where:
$$r_1 : e_{1,1} = [7 - 15] \wedge e_{2,1} \rightarrow e_{3,1}$$
$$r_2 : e_{1,2} = [20 - 30] \wedge e_{2,2} \rightarrow e_{3,2}$$

First, we build the first branch $b_1$ (representing $r_1$) of the tree; this latter has the following format: $[A_{1,1}-e_{1,1}-A_{2,1}-e_{2,1}-A_{3,1}-e_{3,1}-Null]$ (see step1 in figure 5.a). Next, we consider how to join $b_2$ (representing $r_2$) to the tree. We note that $e_{1,2} \cap e_{1,1} = \phi$. However, for any packet whose value of attribute $A_1$ is in the set $[7-15]$, it does not match $r_2$. Thus, we proceed as follows:
- We make a new edge in the tree from $A_{1,2}$ ($A_{1,2}=A_{1,1}$) labeled $e_{1,2}=[20-30]$ (see step2 in figure 5.a).
- We build $Suffix\_node(b_2, A_{2,2})$ that we attach to the node $e_{1,2}$ (see step3 in figure 5.a).
- We update the decision tree structure notation.

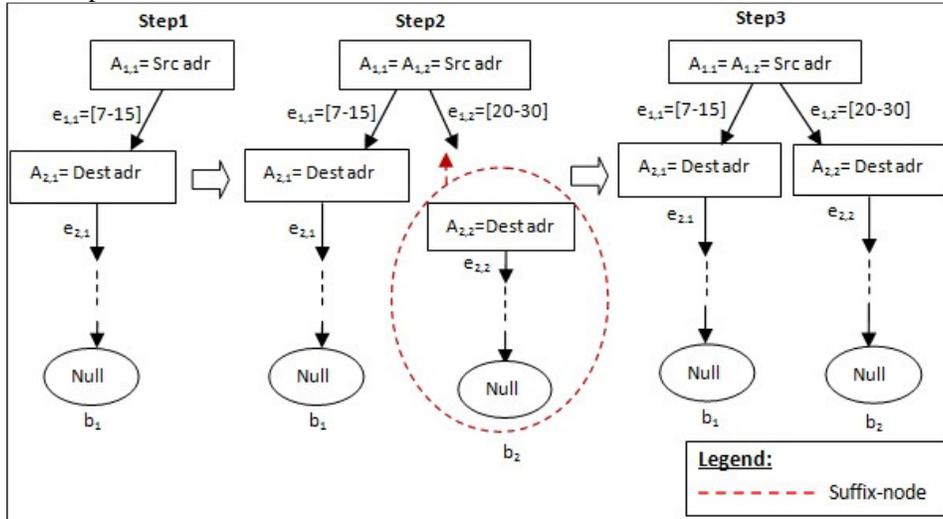

Figure 5.a Case 1: $r_1$ $R_D$ $r_2$

**-Case 2**: $r_1 \Re_{IM} r_2$ : Let's take, for example, the set of 2 rules $R_x \{r_1, r_2\}$ where:

$$r_1 : e_{1,1} = [7 - 15] \wedge e_{2,1} \rightarrow e_{3,1}$$
$$r_2 : e_{1,2} = [7 - 20] \wedge e_{2,2} \rightarrow e_{3,2}$$

We first build the first branch $b_1$ (representing $r_1$) of the tree; this latter has the following format: $[A_{1,1}$-$e_{1,1}$-$A_{2,1}$-$e_{2,1}$-$A_{3,1}$-$e_{3,1}$-$Null]$ (see step1 in figure 5.b). Next, we consider how to join $b_2$ (representing $r_2$) to the tree. We note that $e_{1,1} \subset e_{1,2}$. However, for any packet whose value of attribute is in the set [7-15], it may match the first rule $r_1$, and it also may match $r_2$. Thus,

- We make a new edge in the tree from $A_{1,2}$ ($A_{1,2}=A_{1,1}$) labeled $e_{1,2} = [e_{1,2} - e_{1,1}]$.
- We build $Suffix\_node(b_2,A_{2,2})$ that we attach to the edge $e_{1,2}$ (see step2 in figure 5.b).
- We attach $Suffix\_edge(b_2,e_{2,2})$ to the node to which $e_{1,1}$ points to(see step3 in figure 5.b).
- We update the decision tree structure notation.

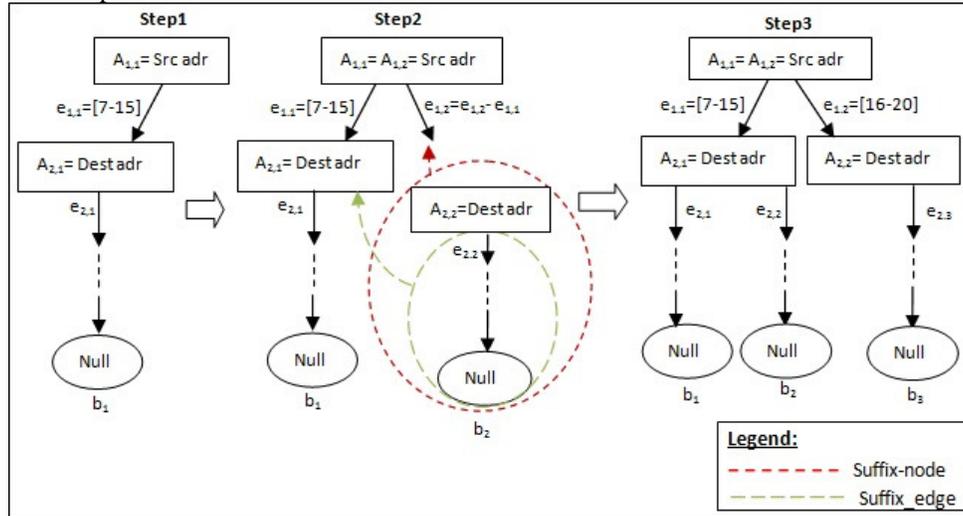

Figure 5.b Case 2: $r_1$ $R_{IM}$ $r_2$

**-Case 3:** $r_2 \mathfrak{R}_{IM} r_1$ : Let's take, for example, the set of 2 rules $R_x$ {$r_1$, $r_2$} where:

$$r_1 : e_{1,1} = [7 - 15] \wedge e_{2,1} \rightarrow e_{3,1}$$
$$r_2 : e_{1,2} = [8 - 12] \wedge e_{2,2} \rightarrow e_{3,2}$$

We first build the first branch $b_1$ (representing $r_1$) of the tree; this latter has the following format: $[A_{1,1}$-$e_{1,1}$-$A_{2,1}$-$e_{2,1}$-$A_{3,1}$-$e_{3,1}$-$Null]$ (see step1 in figure 5.c). Next, we consider how to join $b_2$ (representing $r_2$) to the tree. We note that $e_{1,2} \subset e_{1,1}$. However, for any packet whose value of attribute is in the set [8-12], it matches $r_1$. Thus, we proceed as follows:

- We make a new edge in the tree from $A_{1,2}$ ($A_{1,2}=A_{1,1}$) labeled $e_{1,2} = [e_{1,1} \cap e_{1,2}]$
- We build $Suffix\_node(b_2,A_{2,2})$ that we attach to the node $e_{1,2}$ (see step2 in figure 5.c).
- We attach $Suffix\_edge(b_1,A_{2,1})$ to the node to which $e_{1,2}$ points to(see step3 in figure 5.c).
- The edge $e_{1,1}$ will be renamed $e_{1,1} = [e_{1,1} - e_{1,2}]$.
- We update the decision tree structure notation.

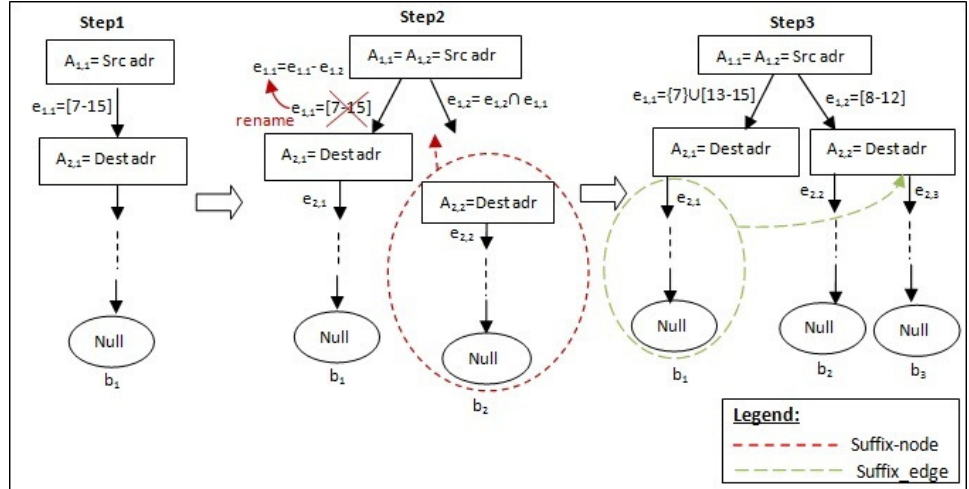

Figure 5.c Case 3: $r_2$ $R_{IM}$ $r_1$

**-Case 4:** $r_1 \Re_c r_2$ : Let's take, for example, the set of 2 rules $R_x$ {$r_1$, $r_2$} where:

$$r_1 : e_{1,1} = [7 - 15] \wedge e_{2,1} \rightarrow e_{3,1}$$
$$r_2 : e_{1,2} = [10 - 20] \wedge e_{2,2} \rightarrow e_{3,2}$$

We first build the first branch $b_1$ (representing $r_1$) of the tree; this latter has the following format: $[A_{1,1}$-$e_{1,1}$-$A_{2,1}$-$e_{2,1}$-$A_{3,1}$-$e_{3,1}$-$Null]$ (see step1 in figure 5.d). Next, we consider how to join $b_2$ (representing $r_2$) to the tree. We note that $e_{1,2} \not\subset e_{1,1}$ and $e_{1,1} \not\subset e_{1,2}$. However, for any packet whose value of attribute is in the set *[10-20]*, it matches values of a sub-set in the set *[7-15]* . As long as, any packet whose value of attribute is in the set *[7-15]*, it matches values of a sub-set in the set *[10-20]*. Thus, we proceed as follows:

- We make 2 new edges in the tree; the first one from $A_{1,2}$ ($A_{1,2}=A_{1,1}$) labeled $e_{1,2} = [e_{1,2} - e_{1,1}]$. The second one from $A_{1,3}$ ($A_{1,3}=A_{1,2}=A_{1,1}$) labeled $e_{1,3} = [e_{1,2} \cap e_{1,1}]$ .
- We build *Suffix_node($b_2$,$A_{2,2}$)* that we attach on one hand, to the new edge $e_{1,2}$, and on the other hand, to the edge $e_{1,3}$ .
- We join *Suffix_edge($b_2$, $e_{2,2}$)* to the node to which $e_{1,1}$ points to(see step2 in figure 5.d). We attach *Suffix_edge($b_1$,$e_{2,1}$)* to the node to which $e_{1,3}$ point to.
- The edge $e_{1,1}$ will be renamed $e_{1,1} = [e_{1,1} - e_{1,2}]$ (see step3 in figure 5.d).
- We update the decision tree structure notation.

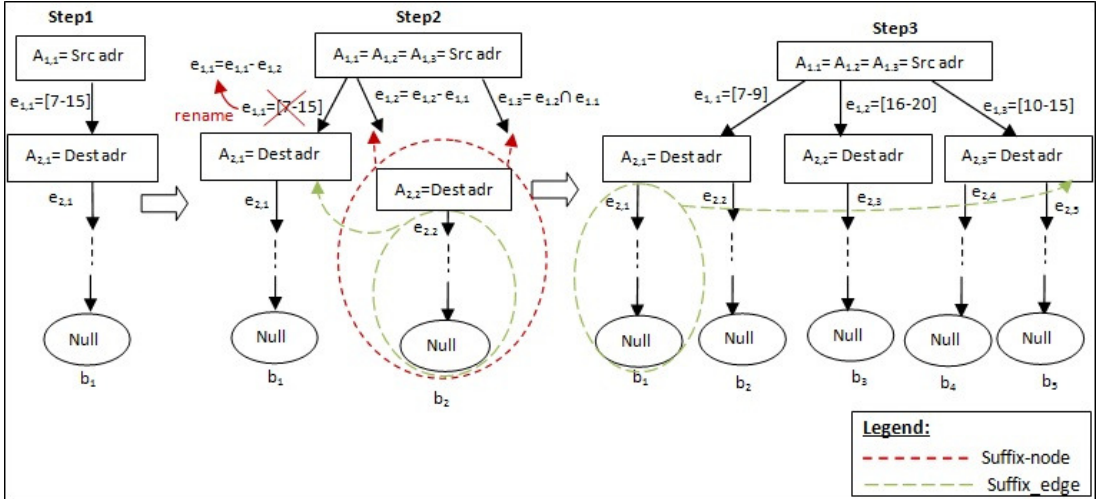

Figure 5.d Case 4: $r_1 R_C r_2$

**-Case 5:** $r_1 \Re_{EM} r_2$: Let's take, for example, the set of 2 rules $R_x \{r_1, r_2\}$ where:

$$r_1 : e_{1,1} = [7 - 15] \wedge e_{2,1} \rightarrow e_{3,1}$$
$$r_2 : e_{1,2} = [7 - 15] \wedge e_{2,2} \rightarrow e_{3,2}$$

We first build the first branch $b_1$ (representing $r_1$) of the tree; this latter has the following format: $[A_{1,1}-e_{1,1}-A_{2,1}-e_{2,1}-A_{3,1}-e_{3,1}-Null]$ (see step1 in figure 5.e). Next, we consider how to join $b_2$ (representing $r_2$) to the tree. We note that $e_{1,2} = e_{1,1}$. However, the two branches share the same edge value. In this case,
- We skip this node $A_{1,1}$ and look for the node $A_{2,1}$ (see step2 in figure 5.e).
- According to the several cases presented above (see cases 1,2,3 and 4), we attach $Suffix\_edge(b_2,e_{2,2})$ to $A_{2,1}$ (see step3 in figure 5.e).
- We update the decision tree structure notation.

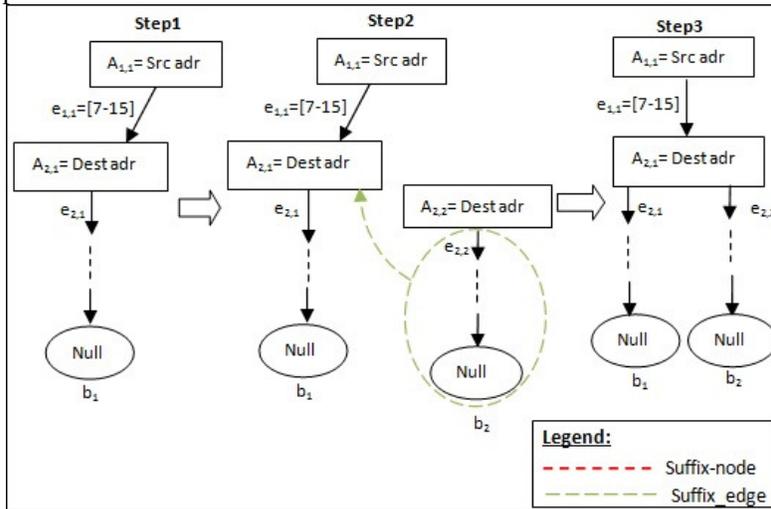

Figure 5.e Case 4: $r_1 R_{EM} r_2$

### 3.3.3 Case Study: Security Component Set Of Rules Correction

In this section, we apply the *RDT* construction principle on the firewall *FW* set of rules in figure 3. Figure 6 illustrates $RDT_{FW}$ the relevant decision tree of the set of rules $R_{FW}$.

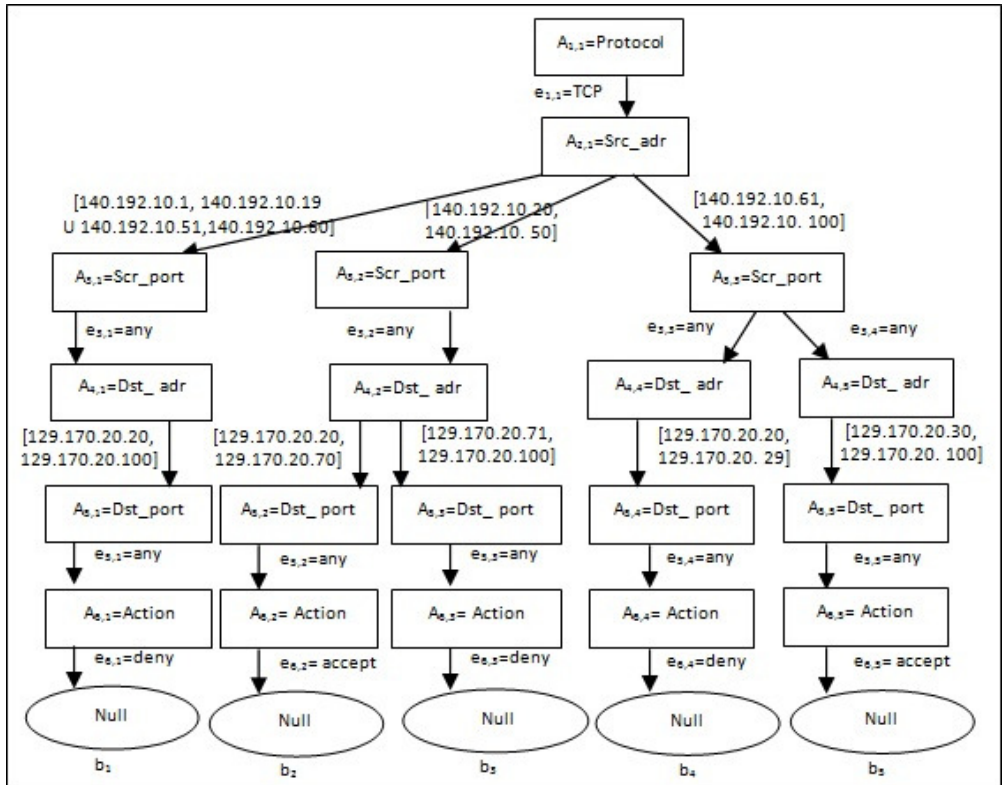

Figure 6. $RDT_{FW}$: The firewall $FW$ relevant decision tree

Now, we convert the $RDT_{FW}$ branches into a set of rules. Based on Lemma 1, we note that these rules are free of anomalies (see table 1).

| Rules | Prtcl | Source address | Src. port | Destination address | Dest. port | Action |
|---|---|---|---|---|---|---|
| $r_1$ | TCP | [140.192.10.1, 140.192.10.19] U [140.192.10.51, 140.192.10.60] | any | [129.170.20.20, 129.170.20.100] | any | deny |
| $r_2$ | TCP | [140.192.10.20, 140.192.10.50] | any | [129.170.20.20, 129.170.20.70] | any | accept |
| $r_3$ | TCP | [140.192.10.20, 140.192.10.50] | any | [129.170.20.71, 129.170.20.100] | any | deny |
| $r_4$ | TCP | [140.192.10.61, 140.192.10.100] | any | [129.170.20.20, 129.170.20.29] | any | deny |
| $r_5$ | TCP | [140.192.10.61, 140.192.10.100] | any | [129.170.20.30, 129.170.20.100] | any | accept |

Table 1. The firewall $FW$ set of relevant rules

## 4. Security Components Interoperability Checking (Process 2)

Let's take a distributed network composed of two relevants security components: the firewall "*FW*" and an intrusion detection system "*IDS*". Now, we will study *FW* and *IDS* interoperability in the network. To do that, we will study if there are misconfigurations between them.

### 4.1 Step D: Security Components Set Of Attributes Extraction (See Figure 1)

Let's suppose that *FW* and *IDS* are composed, respectively, of the set of rules $R_{FW}$ and $R_{IDS}$. $R_{FW}$ is a set of *t* rules $\{r_1, r_2, \ldots r_i, \ldots r_t\}$ where *i* is the relative position of a rule within $R_{FW}$. As far as, $R_{IDS}$ is a set of *z* rules $\{q_1, q_2, \ldots q_j, \ldots q_z\}$ where *j* is the relative position of a rule within $R_{IDS}$.

Each rule $r_i$ belonging to $R_{FW}$ has the following attributes: $Att_{FW}=\{$ *Protocol, Source address, Destination address, Source port, Destination port*$\}$ (see table 1). In the same way, each rule $q_j$ belonging to $R_{IDS}$ has the following attributes: $Att_{IDS}= \{$*Packet length, Protocol, Source address, Destination address, Source port, Destination port, Attack class*$\}$. Table 2 presents the *IDS* set of

International Journal of Computer Networks & Communications (IJCNC) Vol.5, No.5, September 2013

rules $R_{IDS}$. Based in table 1 and table 2, we note that the two security components *FW* and *IDS* differ in attributes number.

| Rules | Packet length | Prtcl | Source address | Src. port | Destination address | Dest. port | Attack Class | Action |
|---|---|---|---|---|---|---|---|---|
| $r_1$ | All | TCP | [140.192.10.40, 140.192.10.50] | any | [129.170.20.10, 129.170.20.70] | any | winworm | reject |
| $r_2$ | All | TCP | [140.192.10.70, 140.192.10.90] | any | [129.170.20.30, 129.170.20.50] | any | winworm | reject |
| $r_3$ | 10 | UDP | 140.192.20.* | any | 210.160.20.* | any | Win32 | reject |

Table 2. The intrusion detection system *IDS* set of relevants rules

### 4.2 Step E: Security components set of rules extension (see figure 1)

To be able to check *FW* and *IDS* interoperability in a network, they must share the same attributes. For that, we will extend the firewall *FW* set of rules format by adding the complementary attributes from the intrusion detection system *IDS* set of rules format and vice versa. The extended rules format, taking into account *FW* and *IDS* attributes is the following:

$Att_{FW} \cup Att_{IDS} = \{$ Packet length, Protocol, Src. address, Src port, Dest. Address, Dest. Port, Attack Class, Action $\}$

Applying the extended format to *FW* set of rules, we obtain the following extended set of rules (see table 3). We note that for each attribute which has not a specific value, we put in the corresponding field "All". "All" means that this field accepts any value defined in the attribute's domain. The intrusion detection system *IDS* set of rules remains unchanged seeing that its set of attributes are conform to the extended rule format.

| Rules | Packet length | Prtcl | Source address | Src. port | Destination address | Dest. port | Attack Class | Action |
|---|---|---|---|---|---|---|---|---|
| $r_1$ | All | TCP | [140.192.10.1, 140.192.10.19] U [140.192.10.51, 140.192.10.60] | any | [129.170.20.20, 129.170.20.100] | any | All | deny |
| $r_2$ | All | TCP | [140.192.10.20, 140.192.10.50] | any | [129.170.20.20, 129.170.20.70] | any | All | accept |
| $r_3$ | All | TCP | [140.192.10.20, 140.192.10.50] | any | [129.170.20.71, 129.170.20.100] | any | All | deny |
| $r_4$ | All | TCP | [140.192.10.61, 140.192.10.100] | any | [129.170.20.20, 129.170.20.29] | any | All | deny |
| $r_5$ | All | TCP | [140.192.10.61, 140.192.10.100] | any | [129.170.20.30, 129.170.20.100] | any | All | accept |

Table 3. The firewall *FW* set of extended rules

### 4.3 Step F: Formal Security Components Interoperability Checking (See Figure 1)

Several works [13,14,15] have defined a set of anomalies detectable between rules in distributed security components called "distributed component anomalies". In the following, we will study these anomalies using the decision tree formalism. Then, we will propose a formal method to remove them.

**Definition 5:**
Let's take a network composed of a set of distributed hosts and several security components. Let a traffic stream flowing from sub-domain $Dom_x$ to sub-domain $Dom_y$ across two security components $C_x$ and $C_y$ installed on the network path between the two sub-domains (see figure 7) [14,15]. At any point on this path in the direction of flow, $C_x$ is called the *preceeding security component* whereas $C_y$ is called a *following security component*.

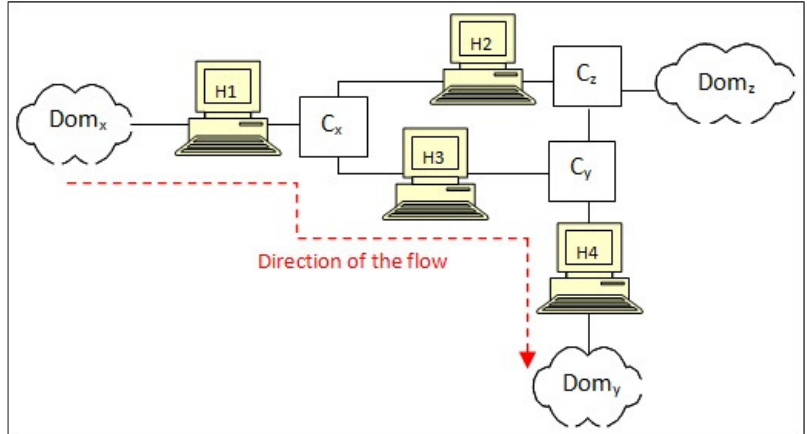

Figure 7. The distributed architecture

### 4.3.1 Distributed Security Components Anomalies Detection

In this section, we classify anomalies that may exist between rules in multi-security component environments. Let's take a rule $r_i$ ($1 \leq i \leq t$) belonging to the preceding security component $C_x$ set of rules $D_x$, and a rule $q_j$ ($1 \leq j \leq z$) belonging to the following security component $C_y$ set of rules $D_y$. We assume that every security component is relevant.

**Property 8: Inter-Shadowing Anomaly**

Let's take two security components $C_x$ and $C_y$. A shadowing anomaly occurs if the preceding security component $C_x$ blocks the network traffic accepted by the following security component $C_y$. In the decision tree representation, for any two branches $b_i$ and $b_j$ belonging respectively to $RDT_{Cx}$ and $RDT_{Cy}$, $b_j$ is shadowed by $b_i$ if and only if,

$$(b_j \in RDT_{Cy}) \, \Re_{IM} \, (b_i \in RDT_{Cx}) \wedge (e_{n,i} = deny/reject) \wedge (e_{n,j} = accept/pass)$$

**Property 9: Inter-Spuriousness Anomaly**

Let's take two security components $C_x$ and $C_y$. A spuriousness anomaly (also called misconnection anomaly) occurs if the preceding security component $C_x$ permits the network traffic denied by the following security component $C_y$. In the decision tree representation, for any two branches $b_i$ and $b_j$ belonging respectively to $RDT_{Cx}$ and $RDT_{Cy}$, $b_i$ allows a spurious traffic to $b_j$ if and only if,

$$(b_j \in RDT_{Cy}) \, \Re_{IM} \, (b_i \in RDT_{Cx}) \wedge (e_{n,i} = accept/pass) \wedge (e_{n,j} = deny/reject)$$

**Property 10: Inter-Redundancy Anomaly**

Let's take two security components $C_x$ and $C_y$. A redundancy anomaly occurs if the following component $C_y$ denies the network traffic already blocked by an preceding component $C_x$. In the decision tree representation, for any two branches $b_i$ and $b_j$ belonging respectively to $RDT_{Cx}$ and $RDT_{Cy}$, $b_j$ is redundant to $b_i$ if and only if,

$$(b_i \in RDT_{Cx}) \, \Re_{IM} \, (b_j \in RDT_{Cy}) \wedge (e_{n,i} = deny/reject) \wedge (e_{n,j} = deny/reject)$$

**Property 11: Inter-Correlation Anomaly**

Let's take two security components $C_x$ and $C_y$. A correlation anomaly occurs as a result of having two correlated rules in the preceding and following components. As defined in section 3.2.2, a security component has a correlated rules only if these rules have different filtering actions. However, correlated rules having any action are always a source of anomaly in distributed components because of the implied rule resulting from the conjunction of the correlated rules. This creates not only ambiguity in the inter-components set of rules, but also spurious, and

111

shadowing anomalies. In the decision tree representation, for any two branches $b_i$ and $b_j$ belonging respectively to $RDT_{Cx}$ and $RDT_{Cy}$, $b_j$ and $b_i$ are correlated if and only if,

$$(b_i \in RDT_{Cx}) \Re_C (b_j \in RDT_{Cy}) \land (e_{n,i} = accept/pass) \land (e_{n,j} = deny/reject)$$

or

$$(b_i \in RDT_{Cx}) \Re_C (b_j \in RDT_{Cy}) \land (e_{n,i} = deny/reject) \land (e_{n,j} = accept/pass)$$

**Property 12: Interoperability**

Security components in a distributed system are interoperable, if and only if, for any two security components $(C_x, C_y)$ where $C_x$ is the preceding security component and $C_y$ is the following security component, there are no anomalies between them (inter-shadowing anomaly, inter-spuriousness anomaly, inter-redundancy anomaly and inter-correlation anomaly)

### 4.3.2 Case study: Distributed Security Components Anomalies Detection

In our case study, based on table 2 and table 3, we note that the firewall *FW* and the intrusion detection system *IDS* contain some misconfigurations between their rules:

- $b_5$ in $RDT_{FW}$ (representing $r_5$ in *FW*) allows a spurious traffic to $b_2$ in $RDT_{IDS}$ (representing $r_2$ in *IDS*). More precisely:

$$\left[(e_{1,5} \in RDT_{FW}) = (e_{1,2} \in RDT_{IDS})\right] \land \left[(e_{2,5} \in RDT_{FW}) = (e_{2,2} \in RDT_{IDS})\right] \land \left[(e_{3,5} \in RDT_{FW}) \supset (e_{3,2} \in RDT_{IDS})\right] \land$$

$$\left[(e_{4,5} \in RDT_{FW}) = (e_{4,2} \in RDT_{IDS})\right] \land \left[(e_{5,5} \in RDT_{FW}) \supset e_{5,2} \in RDT_{IDS})\right] \land \left[(e_{6,5} \in RDT_{FW} = e_{6,2} \in RDT_{IDS})\right] \land$$

$$\left[(e_{7,5} \in RDT_{FW}) \supset (e_{7,2} \in RDT_{IDS})\right] \land \left[(e_{8,5} \in RDT_{FW}) \neq (e_{8,2} \in RDT_{IDS})\right]$$

- $b_2$ in $RDT_{FW}$ (representing $r_2$ in *FW*) is correlated with $b_1$ in $RDT_{IDS}$ (representing $r_1$ in *IDS*). More precisely:

$$\left[(e_{1,2} \in RDT_{FW}) = (e_{1,1} \in RDT_{IDS})\right] \land \left[(e_{2,2} \in RDT_{FW}) = (e_{2,1} \in RDT_{IDS})\right] \land \left[(e_{3,2} \in RDT_{FW}) \supset (e_{3,1} \in RDT_{IDS})\right] \land$$

$$\left[(e_{4,2} \in RDT_{FW}) = (e_{4,1} \in RDT_{IDS})\right] \land \left[(e_{5,2} \in RDT_{FW}) \subset (e_{5,1} \in RDT_{IDS})\right] \land \left[(e_{6,2} \in RDT_{FW}) = (e_{6,1} \in RDT_{IDS})\right] \land$$

$$\left[(e_{7,2} \in RDT_{FW}) \supset (e_{7,1} \in RDT_{IDS})\right] \land \left[(e_{8,2} \in RDT_{FW}) \neq (e_{8,1} \in RDT_{IDS})\right]$$

Thus, they don't verify property 12. Therefore, they are non-interoperable in the network. In the next section, we will present a novel approach to remove these conflicts in order to guarantee their perfect interoperability between *FW* and *IDS*.(see process 3 in figure 1).

## 5. Security Components Interoperability Correction (Process 3)

The interoperability correction process guarantees the perfect interoperability between security components in a network. It is composed of the followings steps:

### 5.1 Step G: Security Components Set Of Rules Integration (See Figure 1)

In this step, we will put together the two security components set of rules in order to detect and correct misconfigurations between them (See step G in figure 1). For that, considering that the firewall is the preceding security component and the intrusion detection system is the following security component, we add *IDS* set of rules to those of *FW*. Eventually, we will update *IDS* set of rules order to get a coherent global set of rules (see column "Rules" in table 4).

International Journal of Computer Networks & Communications (IJCNC) Vol.5, No.5, September 2013

| FW set of rules | | | | | | | | |
|---|---|---|---|---|---|---|---|---|
| Rules | Packet length | Prtcl | Source address | Src. port | Destination address | Dest. port | Attack Class | Action |
| $r_1$ | All | TCP | [140.192.10.1, 140.192.10.19] U [140.192.10.51,140.192.10.60] | any | [129.170.20.20, 129.170.20.100] | any | All | deny |
| $r_2$ | All | TCP | [140.192.10.20, 140.192.10.50] | any | [129.170.20.20, 129.170.20.70] | any | All | accept |
| $r_3$ | All | TCP | [140.192.10.20, 140.192.10.50] | any | [129.170.20.71, 129.170.20.100] | any | All | deny |
| $r_4$ | All | TCP | [140.192.10.61, 140.192.10.100] | any | [129.170.20.20, 129.170.20.29] | any | All | deny |
| $r_5$ | All | TCP | [140.192.10.61, 140.192.10.100] | any | [129.170.20.30, 129.170.20.100] | any | All | accept |

+

| ID set of rules | | | | | | | | |
|---|---|---|---|---|---|---|---|---|
| Rules | Packet length | Prtcl | Source address | Src. port | Destination address | Dest. port | Attack Class | Action |
| $r_1$ | All | TCP | [140.192.10.40, 140.192.10.50] | any | [129.170.20.10,129.170.20.70] | any | winworm | reject |
| $r_2$ | All | TCP | [140.192.10.70, 140.192.10.90] | any | [129.170.20.30, 129.170.20.50] | any | winworm | reject |
| $r_3$ | 10 | UDP | 140.192.20.* | any | 210.160.20.* | any | Win32 | reject |

⇓

| FW U ID set of rules | | | | | | | | |
|---|---|---|---|---|---|---|---|---|
| Rules | Packet length | Prtcl | Source address | Src. port | Destination address | Dest. port | Attack Class | Action |
| $r_1$ | All | TCP | [140.192.10.1, 140.192.10.19] U [140.192.10.51,140.192.10.60] | any | [129.170.20.20, 129.170.20.100] | any | All | deny |
| $r_2$ | All | TCP | [140.192.10.20, 140.192.10.50] | any | [129.170.20.20, 129.170.20.70] | any | All | accept |
| $r_3$ | All | TCP | [140.192.10.20, 140.192.10.50] | any | [129.170.20.71, 129.170.20.100] | any | All | deny |
| $r_4$ | All | TCP | [140.192.10.61, 140.192.10.100] | any | [129.170.20.20, 129.170.20.29] | any | All | deny |
| $r_5$ | All | TCP | [140.192.10.61, 140.192.10.100] | any | [129.170.20.30, 129.170.20.100] | any | All | accept |
| $r_6$ | All | TCP | [140.192.10.40, 140.192.10.50] | any | [129.170.20.10,129.170.20.70] | any | winworm | reject |
| $r_7$ | All | TCP | [140.192.10.70, 140.192.10.90] | any | [129.170.20.30, 129.170.20.50] | any | winworm | reject |
| $r_8$ | 10 | UDP | 140.192.20.* | any | 210.160.20.* | any | Win32 | reject |

Table 4. The global set of rules

## 5.2 Step H: Formal Global Set Of Rules Correction (See Figure 1)

In this step, we will correct the global set of rules using the relevant decision tree formalism presented above (See section 3.3). The correction step consists in focusing on the set of rules and generating a new one free of anomalies (see example in section 3.3.3).

## 5.3 Step I: Specific Security Component Set Of Rules Extraction From The Global Set Of Rules (See Figure 1)

To get a specific security component set of rules, we must extract it from the global one. From the returned relevant decision tree in step H, (see section 5.2) we will extract a sub-tree which represents the specific security component set of rules. This extraction is based on the specific security component predefined attributes. In the following section, we will define a projection operator which accepts, as input, a set of predefined security component set of attributes and the global security set of rules, and returns, as output, a specific security component set of rules in the form of a decision tree (see step I in figure 1).

### 5.3.1 The Projection Operator

Let's take an $RDT$ composed of $t$ branches and $Att_x=\{A_1,A_2,\ldots,A_n\}$ set of $n$ attributes belonging to a security component $C_x$. In order to extract, from $RDT$, a sub-decision tree $DRT_X$, we define an operator called "component projection" and denoted "$\pi$" as follows:

$$\pi(RDT, Att_X) \rightarrow RDT_X \quad (20)$$

This operator removes all branches in $RDT_X$ whose attributes $A_i$ does not belong to $Att_X$ and their corresponding labeled value $(e_{i,j} \neq All)$.

**Lemma 3**

The component projection $\pi$ preserves the relevancy property.

Applying the projection operator to our case study, we will extract the firewall $FW$ and the intrusion detection system $IDS$ decision trees from the global $RDT_G$. Let $Att_{FW}$ and $Att_{IDS}$ the set of attributes of $FW$ and $IDS$. We note that:

$Att_{FW}=\{$ Prtcl, Src address, Port src, Dest address, Port dest, Action$\}$.
$Att_{ID}= \{$Packet length, Prtcl, Src address, Port src, Dest address, Port dest, Attack class, Action$\}$.

Let $RDT_G$ the relevant decision tree describing the global set of rules returned in step H (see section 5.2). By applying the "component projection", we have the following results:

- For the firewall $FW$, branches $b_6$ and $b_{10}$ will be removed considering that the attribute "Attack class" $\notin Att_{FW}$. Also, the branch $b_{12}$ will be removed considering that the attribute "Packet lenght" $\notin Att_{FW}$.
- For the intrusion detection system $IDS$, we will maintain all branches whose attribute "Attack class" $\neq$ All that are $b_6$, $b_{10}$ and $b_{12}$.

From the returned $RDT_{FW}$, we will remove the labeled edges "All" because these edges are insignificant in the security component's attributes. Contrary to that, $RDT_{IDS}$ remains unchanged considering that the set of used attributes represent $Att_{IDS}$.

Finally, $RDT_{FW}$ and $RDT_{IDS}$ branches will be transformed into a set of rules. Table 5 and table 6 represent $FW$ and $IDS$ set of rules.

| Rules | Prtcl | Src address | Src. port | Dest address | Dest. port | Action |
|---|---|---|---|---|---|---|
| $r_1$ | TCP | [140.192.10.1, 140.192.10.19] U [140.192.10.51,140.192.10.60] | any | [129.170.20.20 , 129.170.20.100] | any | deny |
| $r_2$ | TCP | [140.192.10.20, 140.192.10.39] | any | [129.170.20.20 , 129.170.20.70] | any | accept |
| $r_3$ | TCP | [140.192.10.20, 140.192.10.39] | any | [129.170.20.71 , 129.170.20.100] | any | deny |
| $r_4$ | TCP | [140.192.10.61, 140.192.10.69] U [140.192.10.91,140.192.10.100] | any | [129.170.20.20 , 129.170.20.29] | any | deny |
| $r_5$ | TCP | [140.192.10.61, 140.192.10.69] U [140.192.10.91, 140.192.10.100] | any | [129.170.20.30 , 129.170.20.100] | any | accept |
| $r_6$ | TCP | [140.192.10.40, 140.192.10.50] | any | [129.170.20.20 , 129.170.20.70] | any | accept |
| $r_7$ | TCP | [140.192.10.40, 140.192.10.50] | any | [129.170.20.71 , 129.170.20.100] | any | deny |
| $r_8$ | TCP | [140.192.10.70, 140.192.10.90] | any | [129.170.20.20 , 129.170.20.29] | any | deny |
| $r_9$ | TCP | [140.192.10.70, 140.192.10.90] | any | [129.170.20.51 , 129.170.20.100] | any | accept |

Table 5. The firewall $FW$ set of relevant rules

| Rules | Packet length | Protocol | Source address | Source port | Destination address | Dest. port | Attack Class | Action |
|---|---|---|---|---|---|---|---|---|
| $r_1$ | All | TCP | [140.192.10.40,140.192.10.50] | any | [129.170.20.10,129.170.20.19] | any | winworm | reject |
| $r_2$ | All | TCP | [140.192.10.70,140.192.10.90] | any | [129.170.20.30,129.170.20.50] | any | winworm | reject |
| $r_3$ | 10 | UDP | 140.192.20.* | any | 210.160.20.* | any | Win32 | reject |

Table 6. The intrusion detection system $IDS$ set of relevant rules

# 6. CONCLUSION

In this paper, we have proposed a decision tree based approach to check security components interoperability in a network. The interoperability verification procedure is based on several processes; the first one proceeds with a formal specification, verification and correction of the security component' set of rules. The second process checks the interoperability between several security components in the network. If the interoperability is not confirmed, the third process removes the detected misconfiguration to guarantee the perfect interoperability between the security components in the network. So, our approach ensures, on one hand, the security component consistency and on the other hand, the consistency of the distributed security components in the network.

## REFERENCES

[1] A. Benelbahri, and A. Bouhoula, Tuple Based Approach for Anomalies Detection within Firewall Filtering Rules. ISCC 2007. 12th Vol , Iss , 1-4 pp. 63- 70.
[2] A.X. Liu, Firewall Policy Verification and Troubleshooting, Proc. IEEE Int'l Conf. Comm. (ICC) 2008.
[3] A.X. Liu, and M. Gouda, Complete Redundancy Detection in Firewalls, Proc. 19th Ann. IFIP Conf. Data and Applications Security (2005), pp. 196-209.
[4] M. Gouda, and A.X. Liu, A Model of Stateful Firewalls and Its Properties, Proc. IEEE Int'l Conf. Dependable Systems and Networks (DSN '2005), pp. 320-327.
[5] B. Hari, et al.2000. Detecting and Resolving Packet filter Conflicts. Proceedings of IEEE INFOCOM'00.
[6] H. Hamed, E. Al-Shaer, and W. Marrero, Modeling and Verification of IPsec and VPN Security Policies. Proc. 13th IEEE Int'l Conf. Network Protocols (ICNP '2005), pp. 259-278.
[7] K.Golnabi, R. Min, L. Khan and E. Al-Shaer, : Analysis of Firewall Policy Rule Using Data Mining Techniques, 10th IEEE/IFIP Network Operations and Management Symposium- (NOMS06), April (2006)
[8] P. Eronen, and J. Zitting, 2001. An Expert System for Analyzing Firewall Rules. Proceedings of 6th Nordic Workshop on Secure IT-Systems.
[9] P.Chou, and M. Gray,: On decision trees for pattern recognition, IEEE Symposium on Information Theory, Ann Arbor MI., 69, (1986).
[10] SP. Pornavalai, and T. Chomsiri, Analyzing Firewall Policy with Relational Algebra and its Application. Australian Telecommunication Networks and Applications (ATNAC 2004), Australia.
[11] C. R. P.Hartmann, and P. K.Varshney,: Mehrotra, K. G. and Gerberich, C. L.:Application of information theory to the construction of efficient decision trees, IEEE Trans. Inform. Theory vol. IT-28, No.4, 565-577 (1982).
[12] F. Ben Ftima, K. Karoui and H. Ben Ghezala, A multi-agent framework for anomalies detection on distributed firewalls using data mining techniques. Springer Verlag "Data Mining and Multi-agent Integration", 2009. ISBN: 978-1-4419-0523-9, pp267-278.
[13] F. Ben Ftima, K. Karoui, and H. Ben Ghezala, Misconfigurations Discovery Between Distributed Security Policys Using the Mobile Agent Approach. Proc. ACM "The 11th International Conference on Information Integration and Web-based Applications & Services" (iiWAS2009), Kuala Lampur,Malysia.
[14] E. Al-Shaer, H.Hamed, R. Boutaba, M. Hasan, Conflict classification and analysis of distributed firewall policies, IEEE Journal on Selected Areas in Communications (JSAC) 2005, pp. 2069–2084.
[15] E. Al-Shaer, and H. Hamed, Firewall Policy Advisor for anomaly Detection and Rule Editing, IEEE/IFIP Integrated Management IM'2003.
[16] J. Garcia-Alfaro, F.Cuppens, and N. Cuppens-Boulahia, Analysis of Policy Anomalies on Distributed Network Security Setups. Proceedings of the 11th European Symposium on research in computer security (ESORICS 2006), Hamburg, Germany.
[17] E. Al-Shaer, and H. Hamed, Discovery of Policy Anomalies in Distributed Firewalls, Proc. IEEE INFOCOM '2004, pp. 2605-2615.